\documentclass[twocolumn,aps,prb,showpacs]{revtex4-1}

\usepackage{graphicx}
\usepackage{rotating}
\usepackage{amsmath}
\usepackage{amsfonts}
\usepackage{amssymb}
\usepackage{enumerate}
\usepackage{longtable}
\usepackage{dcolumn}
\usepackage{bm}

\begin{document}

\newcommand{\be}{\begin{equation}}
\newcommand{\ee}{\end{equation}}
\newcommand{\bn}{\begin{eqnarray}}
\newcommand{\en}{\end{eqnarray}}

\author{S. D. Das$^{1,4}$, M. S. Laad$^{2,5}$, L. Craco$^{3}$, J. Gillett$^{1}$, V. Tripathi$^{4}$ and S.E. Sebastian$^{1}$ }
\title{Quantum Criticality in the 122 Iron Pnictide Superconductors Emerging from Orbital-Selective Mottness}
\affiliation{
$^{1}$Cavendish Laboratory, Cambridge University, JJ Thomson Avenue, Cambridge CB3 OHE, United Kingdom\\
$^{2}$ Theory College, Institut Laue-Langevin, 6 Rue J. Horowitz, 38000 Grenoble, France\\
$^{3}$ Instituto de F\'isica, Universidade Federal de Mato Grosso, 78060-900, Cuiab\'a, MT, Brazil\\
$^{4}$ Department of Theoretical Physics, Tata Institute of Fundamental Research, Homi Bhabha Road, Navy Nagar, Mumbai 400005, India\\
$^{5}$ Theoretical Physics Department, The Institute of Mathematical Sciences, Chennai 600113, India
}
\email{mslaad@imsc.res.in}

\begin{abstract}
      The twin issues of the nature of the “normal” state and competing order(s) in the iron arsenides are central
to understanding their unconventional, high-$T_c$ superconductivity. We use a combination of transport anisotropy
measurements on detwinned Sr(Fe$_{1-x}$Co$_{x}$)$_2$As$_{2}$ single crystals and local density approximation plus dynamical
mean field theory (LDA + DMFT) calculations to revisit these issues. The peculiar resistivity anisotropy and its
evolution with $x$ are naturally interpreted in terms of an underlying orbital-selective Mott transition (OSMT) that
gaps out the $d_{xz}$ or $d_{yz}$ states. Further, we use a Landau-Ginzburg approach using LDA + DMFT input to rationalize
a wide range of anomalies seen up to optimal doping, providing strong evidence for secondary electronic nematic
order. These findings suggest that strong dynamical fluctuations linked to a marginal quantum-critical point
associated with this OSMT and a secondary electronic nematic order constitute an intrinsically electronic pairing
mechanism for superconductivity in Fe arsenides. 
\end{abstract}

\pacs{
74.70.-b,
74.25.Ha,
76.60.-k,
74.20.Rp
}

\maketitle

\section{Introduction}
Understanding the normal phase of the iron arsenide (FeAs) superconductors holds the key to unraveling the microscopic mechanism of their superconductivity. 
In theoretical approaches based on a weakly-correlated
view of these materials, itinerant magnetic fluctuations play an important role
in the emergence of spin density wave (SDW) and superconducting (SC) orders in Fe-arsenides~\cite{schmalian}.
The alternative intermediate-to-strong coupling view accords preeminence to quasi-local spin fluctuations associated with Mott physics~\cite{si}.  
A large body of experiments, especially in the extensively studied 122-FeAs 
family, are now poised to constrain theory.
Approaching the antiferromagnetic/superconducting (AFM/SC) boundary from the overdoped and/or high-$T$ side, observation 
of electronic nematic (EN) correlations with an onset significantly above the structural 
($T_{s}$) and antiferromagnetic (AFM) Neel ($T_{N}$) temperatures together with the opposite
sign of resistivity anisotropy to that expected point toward the possibility of orbital-driven EN 
order~\cite{fisher} as the primary order parameter.  Subsequent 
ARPES~\cite{fisher1} and STM~\cite{greene} studies performed significantly above $T_{s},T_{N}$
provide additional support for such a view.  
Careful studies~\cite{uchida,coldea,coldea1} of the 122-FeAs family of superconductors strongly 
suggest that SC peaks at a hitherto enigmatic quantum critical point, identified as the
$T=0$ endpoint of the orthorhombic-tetragonal (O-T) structural transition.  
Corroborating evidence for a quantum critical point
(QCP), tacitly assumed to be an AFM-QCP based on a weak-correlation analysis, is also provided by  
transport and NMR~\cite{sachdev-keimer}, and by electronic Raman scattering~\cite{forget} data. 

  Both orbital-driven and spin-driven nematic 
orders have been proposed as the primary order parameter of the normal phase.
  However, distinguishing between these views in FeAs systems is complicated by
 the close proximity of structural (related to $T_{o},T_{s}$) and
AFM transitions ($T_{N}$): they either occur simultaneously~\cite{structure} or 
very close together, with $T_{s}\geq T_{N}$ (thus AFM always occurs in the
orthorhombic phase).  It is impossible to distinguish 
between them on purely symmetry grounds: both views involve breaking of the 
same (discrete rotational) $C_{4v}$ symmetry.  While more tests are needed 
to resolve this issue, a microscopic approach can be very useful here.

Motivated thus, we undertake a joint theoretical-experimental study to illuminate these issues.  We adopt a strong-correlation 
perspective constrained by the above observations. We perform dynamical mean field theory (DMFT) calculations on a multiband Hubbard model 
with first-principles bandstructures for the 122-iron arsenides. Experimentally, we focus on transport anisotropy in Sr(Fe$_{1-x}$Co$_{x}$)$_{2}$As$_{2}$ (Sr-Co-122) 
as the system is tuned through a simultaneous structural and magnetic transition at $T(x_{QCP})\rightarrow 0$ (unlike Ba(Fe$_{1-x}$Co$_{x}$)$_{2}$As$_{2}$, where 
structural criticality precedes the AFM one\cite{rongwei,gillett}). Wherever applicable, we also analyze extant results for 
Ba(Fe$_{1-x}$Co$_{x}$)$_{2}$As$_{2}$ in this picture.  Our main finding is an onset of an orbital selective Mott transition (OSMT) near optimal doping, signaled by the
appearance of a pole in the $xz$-orbital self-energy at $E_{F}$. This naturally explains the near-insulating behavior along the orthorhombic $b$ 
direction observed experimentally. The possibility of an OSMT in pnictides has been proposed earlier based on bandstructure calculations\cite{capone} as well as
in the context of iron selenides\cite{yu-si} using a slave-spin technique. To get deeper insight, we construct a Landau-Ginzburg-Wilson model-- a non-analytic functional of the orbital 
nematic order parameter-- for describing the competing interactions between superconductivity and nematic order parameters, and discuss their effects 
on the nature of transitions between different phases in 122-FeAs. 

\section{Experimental Details}
We use single crystals of Sr-Co-122 grown from FeAs self-flux to study the evolution of transport 
anisotropy similar to previous experiments ~\cite{fisher,rongwei}. Crystals of dimensions of approximately 3mm x 4mmx 1 mm were obtained. 
The crystals grow naturally with the flat surface of the crystal perpendicular to the tetragonal c-axis. The crystals  were cleaved, annealed and 
checked for single domain using Laue crystallography. The orthorhombic unit cell is rotated by $45^{o}$ with respect to the tetragonal unit cell. The crystal was 
oriented and  parallel cuts were made perpendicular to the [110] tetragonal direction so that a rectangular shaped crystal was obtained. A clamp capable of 
providing uniaxial pressure was designed for the purpose of detwinning. The clamp was made of hysol and had brass screws with a stainless steel spring 
which provided a force of 20 N when completely compressed, which resulted in a force of 1.75 N per pitch of the screw. This translated to a pressure of 
about 5-10 MPa on the sample which was placed carefully between the front two pieces of the clamp assembly. Four probe contacts were made using 50 micron 
gold wire. Circuitworks CW2400 was used for making the initial contact and was heat treated for better mechanical strength. Epolead 4929 silver epoxy was 
then used for making good electrical connection. The contact resistance was around 2 ohms.  A continuous flow cryostat was used to do the resistance measurements. 
An ac signal source of frequency 77 Hz supplied a current of 50 microAmps and the voltage was measured using a Princeton Instruments E G \& G Lock-In Amplifier. 
Magneto-resistance measurements were  done on electron doped crystals in a $15T$ magnet supplied by Oxford Instruments.
 
\begin{figure}[t]
\includegraphics[width=0.9\linewidth]{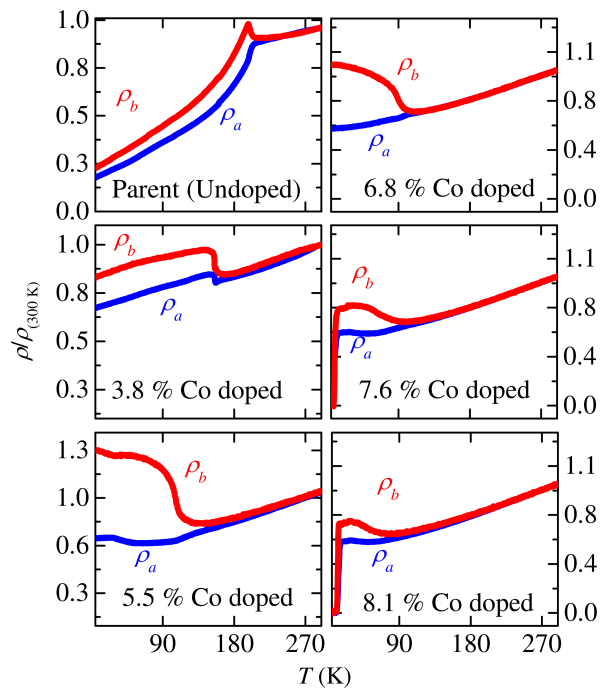} 
\caption{(Color online) Measurements showing the variation of resistivity along the orthogonal $a$ and $b$ directions, with temperature, for various Co dopings. 
For the undoped SrFe$_2$As$_2,$ although the resistivities are different along the two directions below the SDW transition, they are both still metallic. With 
doping the transition is shifted to lower temperatures and vanishes beyond a doping of 8.1 \%, whereas superconductivity appears around 7.6 \%. 
The resistivity along the $b$ direction shows an insulating temperature dependence for intermediate dopings. See text for theoretical analysis.
}
\label{fig1}
\end{figure}

\begin{figure}[]
\includegraphics[width=0.9\linewidth]{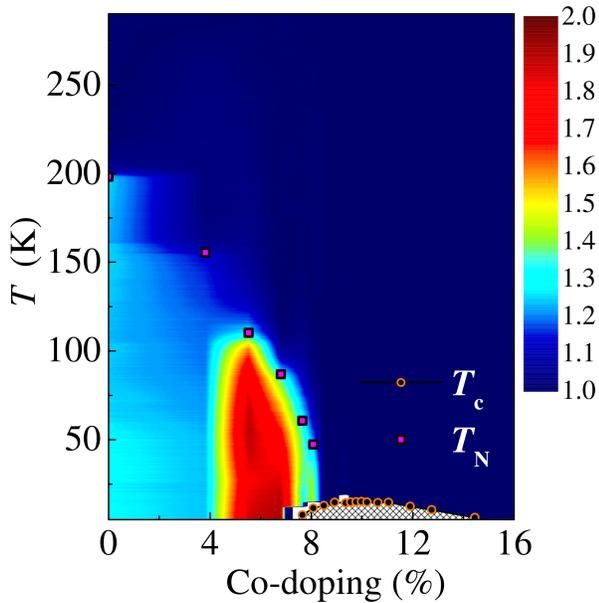} 
\caption{(Color online) Evolution of the anisotropy in resistivity, defined as $[\rho_b/\rho_a ]$ as a function of Co doping in SrFe$_2$As$_2.$ 
The parent compound has a $T_{\text{SDW}} \sim$ 200 K which gradually decreases with increasing electron doping. SC and SDW co-exist for 7.6 \% and 8.1 \% Co doping.~\cite{rongwei}
The anisotropy becomes maximum just prior to the onset of SC and the value is very similar to that obtained in the Ba analogue (see Ref.~\cite{fisher} for a comparison). 
However the splitting between the structural and SDW is not prominent as other cases.
}
\label{fig2}
\end{figure}

Resistivity measurements were done on first on the twinned crystal which measured $ R_{Tw} = \frac{\rho _a + \rho_b}{2}$. Subsequently, pressure 
was applied using the clamps and resistance was measured again which gave $\rho_a$. The anisotropy defined as $(\frac{\rho_b}{\rho_a} -1)$ was obtained 
from these two measurements.
The $T$-dependence of  $\rho_a$ and $\rho_b$ for six Co dopings is shown in
Fig.~\ref{fig1}. From resistivity data we construct a phase diagram (see Fig.~\ref{fig2}), 
where we show the resistivity anisotropy $\rho_b/\rho_a$ as a function of $T$ and doping $x.$ 
%
%

%
Several features stand out: 
$(i)$ the resistivity anisotropy $\delta \rho = (\rho_{b}/\rho_{a})>1$ is in the opposite direction to the lattice anisotropy $b/a - 1<0,$
$(ii)$ $\delta\rho$ evolves non-monotonically, peaking in the vicinity of 6\% doping and disappearing beyond 8\% doping (i.e.) close to
$x_{QCP},$
$(iii)$ $\delta\rho$ acquires a finite value significantly above $T_{s},T_{N}$,
in accord with earlier finding, and 
$(iv)$ anomalous transport (bad-metallic $\rho_{a,b}(T)\simeq 0.025$ ohm-cm at low $T$ even for $x=0$) is 
enhanced with doping, with $\rho_{b}(T,x)$ even showing insulator-like $T$-dependence below $120-150$~K, depending on $x$, while $\rho_{a}(T)$ 
exhibits enhanced bad-metallicity.  This trend is quite intriguing, as a naive expectation mandates enhanced (``good'') metallicity upon doping.

Taken together with earlier data for Ba(Fe$_{1-x}$Co$_{x}$)$_{2}$As$_{2}$, the intriguing features are indicative of an unusual QCP involving electronic 
nematic (EN) order, situated around $x_{QCP}$ where SC is maximized.  However, the insulating temperature-dependence of $\rho_{b}$ also indicates an orbital-selective Mott
transition (of electronic states contributing to $\rho_{b}$) at work, and the loss of both these features 
in the overdoped regime suggests localization physics in a possible EN QCP scenario.  We posit that this localization is not a disorder effect but rather an indicator of an
OSMT, a view supported by recent analysis that finds transport anisotropy to be an intrinsic feature of the renormalized electronic structure~\cite{fisher2}.  
In addition, criticality associated with divergence of the nematic susceptibility
shows ``clean'' critical exponents~\cite{analytis}, suggesting irrelevance of disorder.  While a correct sign of $\delta\rho$ 
can apparently be rationalized within both spin-nematic~\cite{joerg} and ferro-orbital EN~\cite{laad} scenarios, the insulator-like behavior
of $\rho_{b}(T)$ over a wide range of $x$ can be very naturally understood from an OSMT view.
Given that a unified view of the data suggests an intrinsic 
band- (orbital) selective localization tendency at work~\cite{yu-si,laad,kotliar}, we consider this possibility in more detail as well as 
its ramifications for an EN QCP scenario in 122-FeAs systems.  We note that the even more strongly correlated features visible in 
FeSe$_{1-x}$Te$_{x}$~\cite{loidl-B} also make it a good candidate for our proposal. 

\section{Theoretical microscopic treatment} 
To substantiate the link between OSMT and an electronic nematic QCP, 
we have performed first-principles local density approximation plus dynamical mean field theory (LDA+DMFT) 
calculations following earlier work by two of us~\cite{laad}. The
five-$d$ bands of Fe, computed by the linear muffin tin orbital (LMTO) method, were used as inputs in a multi-orbital DMFT formalism.  The multi-orbital 
iterated perturbation theory (MO-IPT) was used as an impurity solver in DMFT: though not exact, it is a computationally fast 
and effective solver, and has been shown to work quantitatively in a variety of contexts~\cite{laad-v2o3,japan,thermo}.  
We chose $U=4.5$~eV, $J_{H}=0.7$~eV and $U'\simeq (U-2J_{H})$ as interaction parameters for the 5-band Hubbard model,
 in accord with values extracted from screened LDA+GW estimates~\cite{kotliar}. As in earlier work~\cite{laad1}, 
ferro-orbital order (FOO) and EN arise via residual {\it intersite} and inter-orbital two-particle interactions in the 
incoherent ``normal'' state found in DMFT calculations. 

\begin{figure}[]
\includegraphics[width=0.85\linewidth]{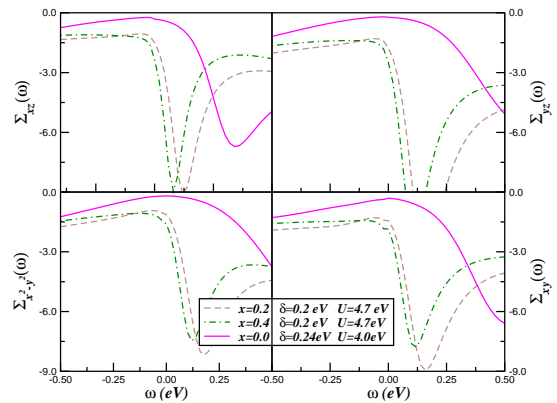} 
\caption{(Color online) Variation of the imaginary part of the self energy corresponding to the $d_{xz}$,$d_{yz}$,$d_{xy}$, and $d_{x^2- y^2}$ orbitals with energy. 
$x$ is the theoretical doping and $(\delta)$ refers to the energy associated with the splitting of the $d_{xz}$ and the $d_{yz}$ bands. As is evident, 
the self energy for the $d_{xz}$ band develops a sharp negative pole near Fermi energy, though slightly shifted implying the onset of the Mott localization.
}
\label{fig3}
\end{figure}

\begin{figure}[]
\includegraphics[width=0.8\linewidth]{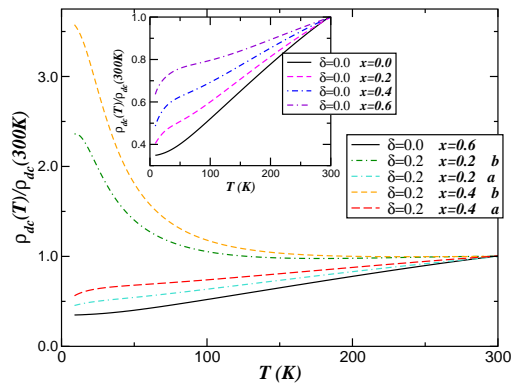} 
\caption{(Color online) Variation of the resistivity along the two orthogonal $a$ and $b$ directions with temperature calculated from DMFT. 
For the parent $( x= 0)$ system although there is split in the response both the directions show metallic behavior as is also noticed in experiments. 
The inset shows the resistivity for different values of doping in the absence of the OSMT~\cite{laad}.
}
\label{fig4}
\end{figure}

Fig.~\ref{fig3} and Fig.~\ref{fig4} clearly mark out the OSMT in our LDA+DMFT calculations. 
Im$\Sigma_{a}(\omega)$ with $a=xz,yz$ in Fig.~\ref{fig3} clearly testify to this as a sharp pole in Im$\Sigma_{xz}(\omega=E_{F}).$  
However, Im$\Sigma_{yz}(\omega)$ reveals bad-metallicity for all $x$: an evident fingerprint of the OSMT.
The OSMT directly implies insulator-like resistivity along $b$.  In Fig.~\ref{fig4}, we show the calculated resistivities along $a,b$, 
computed from the full DMFT Green functions at finite-$T$. The $T$ and $x$-dependence of the calculated resistivity is in very good qualitative agreement 
with resistivity data in Figs.~\ref{fig1} and \ref{fig2}.  In particular, $(i)$ at $x=0$, bad metallic 
resistivity at low $T$ and maximal anisotropy at $T_{s}$ persists above $T_{s}$, as found before~\cite{laad1}, 
and $(ii)$ $\delta\rho$ increases at low $T$ as $x$ increases.  $\rho_{b}(T)$ shows insulator features, even
 as $\rho_{a}(T)$ shows enhanced bad-metallic conductivity, in full accord with data, and $(iii)$ $\delta\rho\simeq 0$ for 
$x>x_{c}\simeq 0.1$, where we define $x=(6+n)/5$~\cite{liebsch} reflecting gradual disappearance of the
EN state and transport anisotropy: this compares favorably with $x_{QCP}\simeq 0.11.$ 

\begin{figure}[]
\includegraphics[width=0.9\linewidth]{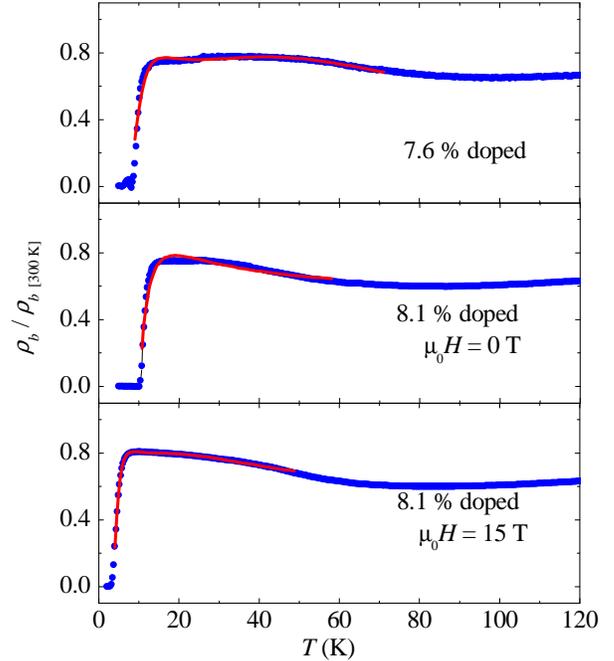} 
\caption{(Color online) 
Halperin-Nelson fits (see text) to the re-
sistivity across the SDW/SC regimes for Co dopings corresponding
to 7.6\% and 8.1\%. Application of magnetic field of 15T lowers the
$T_c$ as can be seen in the bottom panel, and simultaneously increases
the insulating behavior.}
\label{supfig1}
\end{figure}

Additional evidence for an insulating normal state for the $b$ direction and a superconductor-insulator transition driven by phase fluctuations comes from a 
Halperin-Nelson fit of the resitivity data of Fig.~\ref{fig2}. The actual resistivity data near optimal doping is sensitive to superconducting fluctuation effects, and $\rho_b$ does not, at first sight, 
appear to follow an Arrhenius law we expect from a Mott insulator.  This is indeed the case at intermediate $T$, but at lower $T$, this gives 
way to a form characteristic of strong superconducting fluctuations. To separate the effect of superconducting fluctuations, we have fitted the 
resistivity data near $T_c$ (see Fig.~\ref{supfig1})to the well-known Halperin-Nelson interpolation formula, 
$\frac{R}{R_N} = \frac{1}{1 + (\xi/\xi_0)^2},$ where $\frac{\xi}{\xi_0} = a \text{sinh} \frac{b}{\sqrt{t}}$ and $R_N \sim \exp(\Delta/T).$
Here, $t = T - T_0$  while $a,$ $b$ and $\Delta$ are fitting parameters. The Arrhenius behavior in the normal state and the Kosterlitz-Thouless 
like behavior near $T_c$ is typically seen in a superconductor-insulator transition driven by phase fluctuations. 
Together with recent evidence for 
precursor diamagnetism~\cite{canfield} in 122-systems,
this finding further supports selective-Mottness setting in around $x_{opt}$ as a natural contender for possible quantum criticality.  
The key point is that from the number-phase uncertainty principle, an OSMT immediately implies a strong phase 
fluctuation-dominated regime for a subsequent SC instability.  Since both $xz,yz$ states are microscopically implicated in 
the $s_{\pm}$ SC pairing, Mott localization
of (a subset) $xz$ carriers in the ``normal'' state must now necessarily implicate large phase fluctuations above $T_{c}$, 
precisely as indicated by the HN fit.  In addition, we 
also find that an external magnetic field leads to a \textit{positive} magnetoresistance (MR) (which enhances the insulator-like 
$\rho_{b}$ for all $x<x_{QCP}$), while conventional negative MR is recovered for $x>x_{QCP}$, further supporting the 
hypothesis of a QCEP associated with OSMT around $x_{QCP}$.   Without any additional assumptions, OSMT also stabilizes FOO 
and associated EN by enhancing(reducing) $n_{xz}(n_{yz})$ relative to their para-orbital values.
Thus, the EN QCP is intimately tied down to the OSMT, and is thus expected to have an underlying Mott-like criticality. 

\section{Landau-Ginzburg-Wilson phenomenology} 
The existence of the putative electron-nematic QCP, now linked to an OSMT, 
at $x_{QCP}$ corresponding to maximum $T_{c}$ naturally leads us to ask: What role do these coupled criticalities play 
in near-optimally doped Fe-arsenides?  
Since both electronic nematicity and superconducting instabilities result from the same residual interaction, 
how do we describe the competition between the two phases?

  Clearer physical insight into these issues is gained by constructing a Landau-Ginzburg-Wilson (LGW) functional
for the competing ``normal''-SC and ``normal'' metal-EN metal transitions.  
Compared to previous work~\cite{civelli,nevi}, a novel feature of our phenomenology is that LGW parameters for different $x$ are computed 
from the LDA+DMFT incoherent spectral functions (see below) rather than from LDA~\cite{imada}.  
The OSMT in the $xz$ sector corresponds to a pocket-vanishing Lifshitz transition with consequent ferro-orbital 
and electronic nematic instabilities (for which ARPES evidence indeed exists), whence we define the EN order parameter 
$N=\frac{n_{xz}-n_{yz}}{2(n_{yz}+n_{xz})}$, with the free energy expansion~\cite{imada},
$F_{EN}[N]=a N + b_{\mu}N^{2} + c_{\mu}N^{3} + d_{\mu}N^{4}.$
Here, following Yamaji {\it et al.}, the suffix $\mu=p$ refers to the case where there is no selective-Mott transition 
(and associated spontaneous breaking of four-fold rotational symmetry) in the $xz$ orbital 
and $\mu=m$ refers to the symmetry-broken phase ($N < 0$) brought about by the opening of a Mott gap and vanishing of the $xz$ pocket  
across the pocket vanishing Lifshitz point. Since spontaneous symmetry breaking is only on the $N<0$ side, we assume 
$b_p,\,c_p,\,d_p >0.$ 
On the $N<0$ side, $b_m$ can change sign. Thus $F[N]$ \textit{is not} a usual LGW functional since the coefficients are
non-analytic. We incorporate Jahn-Teller orbital-lattice coupling and/or 
uniaxial strain through the renormalization $b_{m}\rightarrow (b_{m}-g^{2}/K)$ (see below).

\begin{figure}
\includegraphics[width=0.9\linewidth]{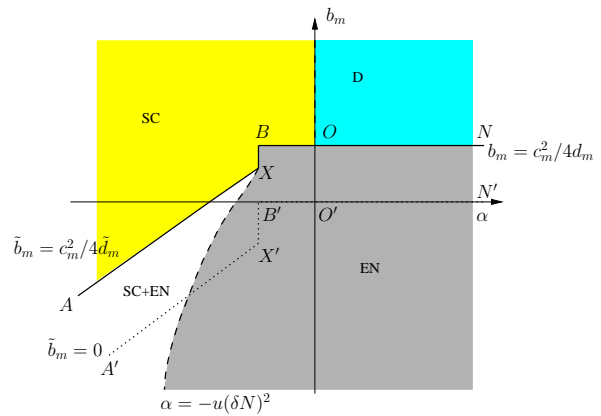} 
\caption{(Color online) Phases on the $b_m$ vs. $\alpha$ plane where $b_m$ is the parameter which changes sign at the Lifshitz QCP 
(in the absence of superconductivity) and $\alpha$ is the coefficient of the quadratic term in the free energy of the superconductor. 
The phase diagram is for the case $c_m > 0$ which is supported by our DMFT calculations.
Here ${\tilde b}_m = b_m - u\alpha/2\beta,$ and ${\tilde d}_m = d_m - u^2/4\beta.$ SC, EN and D refer to superconducting, electron nematic
and disordered phases respectively. The solid lines indicate a first order transition and 
dashed lines indicate second order transitions. The dotted line is the limit of metastability of the non-nematic phases - beyond this line a 
transition to nematic phases (EN or SC+EN) is inevitable. 
Lifshitz transitions correspond to ${\tilde b}_m = 0$ in the coexistence phase and $b_m = 0$ otherwise (line $N'O'B'X'A'$).
}
\label{fig5}
\end{figure}

Introducing now the superconducting (SC) free energy, 
$
F_S[\Psi] = \alpha|\Psi|^2 + \beta |\Psi|^4,
$
and the superconductor-EN coupling 
$F_{NS}[N,\Psi] = u N^2 |\Psi|^2,$
($u>0$) the total free energy is $F = F_{EN}+F_S +F_{NS}.$  
Microscopically, $F_{NS}$ arises from a mean-field decoupling of the inter-site residual interactions with coupled charge-orbital-spin 
character~\cite{laad}.  
For weak coupling of EN and SC order parameters, $u^2 < 4\beta d_m,$ 
the mean-field phase diagram consists of four phases: (i) disordered, $\Psi = N = 0,$ (ii) EN, 
$\Psi = 0,\,N\neq 0,$ (iii) SC, $\Psi \neq 0,\,N=0,$ and (iv) coexisting $\Psi \neq 0,\, N \neq 0.$  
Experimental observation of coexistence of EN and SC phases in underdoped samples would imply a small-$u$ regime in the iron pnictides.
For larger $u$, the coexistence phase is preempted by a first order line separating the EN and SC phases, which is not seen in experiment.
The phase diagram in the $\alpha-b_m$ plane with $c_m > 0,$ (the sign calculated from DMFT) is shown in Fig.~\ref{fig5}.

We briefly sketch our mean field analysis of the coupled EN-superconductor system:
Consider the total free energy $F = F_{EN} + F_{S} + F_{NS},$ where $F_{EN},$ $F_{S}$ and $F_{NS}$ are as described above. 
The mean field solutions $N_{\rm{mf}}$ and $\Psi_{\rm{mf}}$ given by $\delta F/\delta N = 0$ and $\delta F/\delta \Psi^{*} = 0.$  The mean field equations are
\begin{align}
\alpha \Psi + 2\beta |\Psi|^2\Psi + u\Psi N^2 & = 0,
\label{meanfield-sc}\\
a + 2(b_m + u|\Psi|^2)N + 3c_m N^2 + 4 d_m N^3 & = 0,\,\, (N\leq 0).
\label{meanfield-en}
\end{align}
In the following analysis we set the ``Zeeman'' field $a=0$ without loss of generality.
Let us first discuss the case where the superconducting order parameter is zero. Now if $c_m < 0$ (which is the case discussed in Ref.\cite{imada}), we can drop the stabilizing term $d_m N^4.$  For this case, straightforward analysis (see Ref.~\cite{imada}) now shows that a line of first-order transitions 
(for $a<0,b_{m}<0$, where $N$ jumps discontinuously) ends at a {\it marginal} quantum critical end-point (M-QCEP) ($a=0=b_{m}$),
beyond which ($b_m > 0$) only a smooth crossover is obtained. If $c_m > 0,$ the stabilizing term $d_m$ is needed and the phase diagram differs from the $c_m < 0$ case. The key difference is that a first order transition to the EN phase 
is possible even as $b_m > 0$. This transition takes place at 
$b_m = c_m^2/4d_m > 0$ where the compressibility diverges, indicating 
anomalously soft electronic fluctuations associated with the
M-QCEP of the line of first-order transitions associated with EN order. 
The first order transition thus intervenes before the M-QCEP 
($b_m = 0$) can be reached. However, if the first order transition is
weak, proximity to the M-QCEP will still be reflected in the 
physical properties.  In practice, this 
proximity is unearthed by strain-tuning as done by Fisher {\it et al.}

Consider now the coexistence phase, $\Psi\neq 0$ and $N \neq 0.$ We eliminate $|\Psi|^2$ from the free energy using the mean-field solution for
$|\Psi|^2$ and get
\begin{align}
F[N] =  {\tilde b}_m N^2 + c_m N^3 + {\tilde d}_m N^4 - \frac{\alpha^2}{4\beta},
\label{meanfield-coex}
\end{align}
where ${\tilde b}_m = b_m - u\alpha/2\beta$ and ${\tilde d}_m = d_m - u^2/4\beta.$ For stability of this phase, we need ${\tilde d}_m > 0.$
The non-trivial solution for $N_{\rm{mf}} < 0$ is
\begin{equation}
N_{\rm{mf}} = -\frac{3 c_m - \sqrt{9 c_m^2 - 32 {\tilde d}_m{\tilde b}_m}}{8 {\tilde d_m}}. 
\label{N-coex}
\end{equation}
This need not be the solution with the lowest free energy. However once ${\tilde b}_m$ is small enough such that $F[N_{\rm{mf}}] = 0,$ we 
get a first order transition to the coexistence phase. To find where this occurs, we note that we are essentially looking for the condition
for coincident nonzero roots of Eq.\ref{meanfield-coex}. This gives us the condition
\begin{equation}
 c_m^2 = 4{\tilde b}_m{\tilde d_m}
\end{equation}
for the (first order) boundary of the coexistence phase and the disordered phase.

The sign of $c_m$ has a significant effect on the nature of phase transitions in our model. 
For $c_m < 0,$ one would have a line of Lifshitz transitions at $b_m = 0$ in 
the absence of SC and ${\tilde b_m} = b_m - u\alpha/2\beta= 0$ in its presence. 
However, for $c_{m}>0$ that we find in our calculation, a first order transition (line $NOBXA$ in Fig.~\ref{fig5}) 
preempts the Lifshitz transition (line $N'O'B'X'A'$), which is also the limit of metastability. 
The first-order character of the disordered-EN transition drives an 
electronically phase-separated state, that one finds in the case of the pnictides. Electronic 
phase separation is generally expected in the region between the phase transition and limit of metastability 
(see for e.g. Ref.~\cite{chandra}). We find $c_{m}>0(x<x_{L})$ 
and $c_{m}<0(x\geq x_{L})$ from LDA+DMFT calculations ($x_L$ is the value of doping $x$ at which the OSMT takes place), 
which implies that {\it only} underdoped samples 
should exhibit electronic phase separation.  This finding is supported by STM studies on underdoped
CaFe$_{2}$As$_{2-x}$P$_{x}$ \cite{stm-science}.  

It is appropriate to say a few words here about our computation of the coefficients for the free-energy 
functional, $F(N)$, describing the ``normal'' quantum 
paramagnetic (incoherent) metal to the EN phase transition. We have obtained these
from LDA+DMFT calculations on a five-orbital Hubbard model, performed earlier by two of the authors~\cite{laad} 
(see Appendix \ref{A1} for a description). There, a sizably correlated limit of the five-band model was shown to give a good quantitative
accord with a range of one- and two-particle responses in the ``normal'' state
without any symmetry breaking, as well as with key features in both SC~\cite{laad} and (orbital) electronic nematic (EN)~\cite{laad1} states.  
The sizably correlated view is also supported by other first-principles 
approaches~\cite{kotliar} and underlies the frustrated Heisenberg model 
approaches~\cite{phillips-A} to magnetism in Fe arsenides.

In the presence of Jahn-Teller orbital-lattice coupling and/or 
uniaxial strain, the EN-free energy functional is still valid
provided the renormalized value $b_{m}\rightarrow (b_{m}-g^{2}/K)$
is considered in the LGW model. Appendix \ref{A2} contains a discussion
of the renormalization of $b_m$ by orbital-lattice coupling.


We identify $x_{QCP}$ (the doping corresponding to extrapolation of the line of nematic transition to $T=0$) in experiment with $x_L,$ 
the marginal quantum critical endpoint of the first-order line of the OSMT.  Turning to finite-$T$, we 
assume $b_m = b_0(T - T^{*}(x)),$ in the mean-field spirit, where $T^{*}(x) = T'(x_L - x) + O(x_L - x)^2.$ Firstly, this 
implies a divergent {\it charge} nematic susceptibility, $\kappa(x)\simeq b_{m}^{-1}(x,T) = 1/b_0(T - T^{*}(x))$ (i.e,
 the critical exponent, $\gamma=1$) as the Lifshitz point is approached 
from the disordered state.  Such a ``Curie-Weiss`` behavior has indeed been 
observed in strain response~\cite{analytis} and Raman~\cite{forget} data.  
Secondly, the coupling of the nematic order to lattice strain will shift $T_s$ above $T^{*}$ (see Ref.\cite{analytis}). The divergence of 
$\chi_{nem}$ in Raman data at $T^{*}<T_{s}$ can be explained by this picture given that the Raman study unveils
 the strain-free nematic scale.  We note that in this picture, the shift between $T^{*},T_{s}$ is a consequence of coupling of
strain to an {\it intrinsic} EN state (``zeeman'' field on an Ising-like 
orbital nematic), and does not necessarily require that additional (spin 
nematic) mechanisms are needed (Appendix \ref{A3} contains a discussion of strain-nematic coupling and finite-temperature behavior). 
It must be emphasized, though, that onset of 
orbital nematicity will induce a spin-nematic purely on symmetry grounds. While both 
states break the same $C_{4v}$ lattice rotational symmetry of the $T$-phase, an OSMT-induced EN state can naturally account for the data
unlike extant spin-nematic scenarios.

We emphasize that, on pure symmetry grounds, our LGW functional is
entirely consistent with the alternative spin-nematic
view~\cite{schmalian}.  If we were to couple our orbital nematic
order-parameter to the spin-nematic one, we expect to obtain results
comparable to those obtained there, including consistency with the T-x
phase diagram at finite $T$.  However, we have not done this here.
Incorporation of this aspect in a way consistent with an underlying
OSMT in the fermionic theory is left for the future.

A crucially important consequence of our work is its implications
for the stripe-AFM order {\it subsequent} or co-incident with the
structural transition.  Since an OSMT is involved in the (avoided)
quantum criticality,
the Fermi pocket corrsponding to $d_{xz}$ orbital character must
undergo a drastic topological modification.  In a way similar to what
happens in the fractionalized Fermi liquid in the FL$*$
view~\cite{subir}, the $d_{xz}$
hole pocket now corresponds to {\it zeros}, rather than poles of
$G_{xz,xz}(k,\omega)$.  Further since the
$xz-yz$ degeneracy is now non-existent and there are no Landau
quasiparticles in our selective-Mott state, AFM order itself can no
longer arise via conventional nesting instabilities of a bare or
renormalized band structure.

\section{Discussion}
  In summary, our LDA+DMFT+LGW 
analysis explains a broad set of experimental observations and unearths the hitherto unidentified link between soft quasi-local and dualistic electronic fluctuations.
Their combined occurrence at the EN-QCP is identified as a consequence of an underlying marginal quantum critical point at a Lifshitz transition, now associated with an OSMT.

Our theory naturally accounts for a range of additional unusual responses in 
the best-studied 122-FeAs systems.  In the selective-Mott view, vanishing of 
a Fermi surface pocket as the Lifshitz transition is approached from the overdoped 
side corresponds to a
vanishing of an effective coherent bandwidth and a diverging 
effective mass, now associated with the $xz$ carriers, as seen in elastic data.  
An OSMT also naturally rationalizes electrical transport data,  implicating an underlying selective Mott localization in the observed insulating behaviour of $\rho_b.$
Finally, the now orbital dependent 
super-exchanges along $a$ and $b$ also readily lead to a $J_{1a}-J_{1b}-J_{2}$ model via a Kugel-Khomskii mechanism
(now justified ipso facto from selective-Mottness induced local moments) with 
$J_{1a}<J_{1b}$ and $J_{2}>J_{1b}/2$~\cite{phillips-A} {\it and} $J_{1a}<0$.  
This qualitatively accords with the stripe-AF order, as well as the spin-wave spectrum in 
inelastic neutron scattering (INS) studies~\cite{adroja}.


The anomalously soft electronic fluctuations accompanying a 
divergent electronic compressibility close to such a
QCEP emerges as a novel electronic pairing mechanism that can boost superconductivity by
removing the residual entropy of the ``normal'' state incoherent fluctuating liquid.  This scenario is supported by the experimentally accessed
phase diagram, where  nematicity, ferro-orbital order, orthorhombicity and the SDW are suppressed at $T=0$ in close 
proximity to the doping where $T_{sc}(x)$ peaks. 
	
\begin{acknowledgments} 
S.D.D., J.G. \& S.E.S. acknowledge experimental as-
sistance from J. R. Cooper. S.D.D. acknowledges support from Cambridge Commonwealth Trust scholarship. 
M.S.L. thanks the Institut Laue-Langevin and TIFR for financial support. 
L.C.'s work was supported by CAPES - Proc. No. 002/2012. V.T. acknowledges a DST Swarnajayanti grant DST/SJF/PSA-02/2012-2013.  
S.E.S. acknowledges support from the Royal Society, King’s College Cambridge, the Winton Programme for the Physics of 
Sustainability, and the European Research Council grant number FP/2007-2013/ERC Grant Agreement number 337425.

\end{acknowledgments}


\appendix

\section{Free-energy Functional from LDA+DMFT Results}\label{A1}
We start by observing that EN order and associated (local) dynamical 
fluctuations can be incorporated into the DMFT ideology as described 
elsewhere~\cite{laad1}.  The crucial point
to appreciate is that this is tied to ferro-orbital order and the structural 
instability as a result of removal of $d_{xz,yz}$ orbital degeneracy of the 
tetragonal (T) phase.  The upshot is that coupling to a Jahn-Teller or a uniaxial strain term, 
$\lambda\sum_{i}Q_{i}(n_{i,xz}-n_{i,yz})$ (spin indices are suppressed) now
 lowers the 
$d_{xz}$ band by $\lambda\langle Q_{i}\rangle$ and raises the $d_{yz}$ band by 
the same amount.  This mechanism thus offers a simple way of visualising the
relative shift of the $d_{xz,yz}$ bands needed to achieve consistency with the
ARPES FS well above $T_{s},T_{N}$, and it is important to emphasise that it 
is intimately linked to ferro orbital order and resultant (orbital) nematicity. 
It is also precisely this coupling which results in terms odd in $N$
in the LGW free energy used in the main text.

Using the DMFT local spectral functions of the five-band Hubbard model, we
determined the coefficients $d_{L,\pm}^{0}$ and $d_{L,\pm}^{(1)}$ needed to 
compute the co-efficients in $F(N)$ by following Yamaji {\it et al.}~\cite{imada} and linearising the DMFT spectra around $E_{F}$ at and away from the 
Lifshitz point.  Since the EN instability is primarily associated with FOO and 
lifting of the $d_{xz,yz}$ orbital degeneracy, we used only the DMFT results for the $xz,yz$ bands, computed from the full five-orbital problem.  The relevant
formulae are similar to those appearing in Yamaji {\it et al.}~\cite{imada}.

The resulting $a,b_{\mu},c_{\mu}$ are used in the Lifshitz free energy $F_N$ 
and used to derive the main conclusions of the first part in 
this work.  We emphasise that this procedure yields non-analytic co-efficients
used in Eq.(1), and, in contrast to pure phenomenological works, are now 
{\it derived} from the correlated electronic structure (DMFT) input.  In another crucial difference
with other phenomenological approaches, the non-analytic co-efficients in the LGW expansion are a 
non-trivial consequence of the underlying selective-Mott physics found in LDA+DMFT.

\section{Renormalisation of $b_{\mu}$ by orbital-lattice coupling}\label{A2}

We begin with the $d_{xz,yz}$ orbitally degenerate situation in the tetragonal phase.  
As in generic orbital degenerate problems, a symmetry 
adapted Jahn-Teller term will lift this degeneracy, inducing orbital order along with a 
structural distortion to an orthorhombic state.  In Fe arsenides, this ferro-orbital order results in a preferential occupation of the $d_{xz}$ orbital and a
finite orbital nematic order parameter, $N=\frac{n_{xz}-n_{yz}}{2(n_{xz}+n_{yz})}$.  The Jahn-Teller (or orbital-lattice) coupling is

\begin{equation}
H_{J-T}=\lambda\sum_{i}Q_{i}(n_{i,xz}-n_{i,yz})\simeq \lambda\sum_{i}Q_{i}N_{i}
\end{equation}
  In the planar geometry of the FeAs systems, $Q_{i}$ is related to the orthorhombicity, $O=\frac{b-a}{b+a}$ (where $a,b$ are unit cell lattice constants).
In the LGW functional, the orbital-lattice coupling thus induces an extra term
$\lambda\langle Q_{i}\rangle N$, where $\langle Q_{i}\rangle$ is determined by 
minimising 

\begin{equation}
H_{el}=H_{JT}+H_{lat}=\lambda QN + KQ^{2}/2
\end{equation}
with respect to $Q$.  Here, $K$ an effective ``spring constant'' related to 
the details of the phonon spectrum.  This yields $Q=(-\lambda/K)N$, and 
resubstituting this into $H_{J-T}$ yields the extra term $-(\lambda^{2}/K)N^{2}$
which renormalises $b\rightarrow (b-\lambda^{2}/K)$ as stated in the main 
text.  The importance of this effect is seen from the fact that $b$ can now 
change sign, as is needed to derive a line of first-order (now nematic-plus-structural) transitions separated from the region of a smooth crossover by a 
quantum critical end-point where the exotic marginal quantum criticality 
obtains.

\section{Strain-nematic coupling and finite temperature effects}\label{A3}

Here we detail the finite-$T$ GL theory for orbital-nematic order coupled to strain, to simulate
the actual physical situation in experimental studies.  More specifically, we couple the free energy for the 
pocket-vanishing Lifshitz transition,

\begin{equation}
F[N] =\frac{b_{\mu}}{2}N^{2} + \frac{c_{\mu}}{3}N^{3} + \frac{d_{\mu}}{4}N^{4}
\end{equation}
to the strain part, written as

\begin{equation}
F(\epsilon)=\frac{\alpha}{2}\epsilon^{2} + \frac{\beta}{4}\epsilon^{4}
\end{equation}
by a coupling term, $F_{coupl}=-\lambda\epsilon N$.  In the presence of external stress, we follow earlier work~\cite{fisher1}
and differentiate $F=F[N]+F[\epsilon]+F_{coupl}$ with respect to both $N$ and $\epsilon$ to get

\begin{equation}
\frac{d N}{d\epsilon}=\frac{\lambda}{(b_{m}-\lambda^{2}/\alpha)+2c_{m}N+3d_{m}N^{2}}
\end{equation}
Then, in the limit $\lambda\rightarrow 0$ (zero-stress limit), $\delta N\rightarrow 0$, and we get

\begin{equation}
\frac{d N}{d\epsilon}=\frac{\lambda}{b_{m}-\lambda^{2}/\alpha}
\end{equation}
  Thus, in presence of strain the nematic suseptibility diverges at $b_{m}=\lambda^{2}/\alpha$.  Also, above the structural transition, we can 
neglect terms $O(N^{3})$, to get $N=\frac{\lambda}{b_{m}-\lambda^{2}/\alpha}\epsilon$.  Finally, in our mean-field picture
of the transition, we assume that $b_{m}$ has the usual $T$-dependence, i.e, that $b_{m}(T,x)=b_{0}(T-T^{*}(x))$, with $T^{*}(x) \approx T'(x_L - x) +O(x_L - x)^2.$
At $T=0,$ the $x$-dependence comes from LDA+DMFT results as discussed before.

  Thus, the finite-$T$ nematic susceptibility is now
\begin{equation}
\frac{d N}{d\epsilon}=\frac{\lambda}{(b_{m}(T,x)-\lambda^{2}/\alpha)+2c_{m}N+3d_{m}N^{2}}
\end{equation}
whence it follows that finite $N, \epsilon$ occur simultaneously at a {\it renormalized} temperature, $T_{s}=T^{*}+\frac{\lambda^{2}}{b_{0}\alpha} > T^{*}$.  The intrinsic nematic mean-field transition scale is thus {\it lower} than that at which the structural transition occurs
(at $T_{s}$).  Physically, this arises because strain acts as a conjugate field to the nematic order parameter, and so enhances intrinsic 
nematicity.  It also offers a natural explanation of the {\it shift} between the structural transition temperature ($T_{s}$) and the temperature
where resistivity anisotropy is maximum {\it in the normal state}, as seen from the $T-x$ phase diagram.  It is also consistent with the 
fact that, in electronic Raman scattering data~\cite{forget}, the extrapolated charge susceptibility diverges at a $T=T^{*}$ lower than $T_{s}$:
this is now simply because there is no strain effect in Raman studies, which may consequently be 
unearthing the intrinsic $T^{*}=T_{nem}$.  Finally, it is also the reason for apparently different conclusionsin literature, where measurements under strain-tuning find that the EN phase persists into the overdoped
 region in the $T-x$ phase diagram: strain stabilizes EN order but washes out nematic (quantum) 
criticality, while Raman measurements, not carried out under strain, reveal the intrinsic nematic
scale $T_{nem}$.


\end{document}